\documentclass[aps,pra,twocolumn,showpacs,floatfix]{revtex4}
\usepackage{bm}
\usepackage{graphicx} 
\usepackage{amsmath}
\usepackage{amssymb}
\usepackage{epstopdf}
\usepackage{dsfont}
\usepackage{xcolor}

\begin{document}

\title{Ultracold collisions of the lithium monoxide radical}

\author{Lucie D. Augustovi\v{c}ov\'{a}$^{1}$}
\author{John L. Bohn$^{2}$}
\affiliation{$^{1}$Charles University, Faculty of Mathematics and Physics, Department of Chemical Physics and Optics, Ke Karlovu 3, CZ-12116 Prague 2, Czech Republic}

\affiliation{$^{2}$JILA, NIST, and Department of Physics, University of Colorado, Boulder, CO 80309-0440, USA}

\date{\today}

\begin{abstract}

Ultracold collisions of LiO molecules in the $^{2}\Pi_{3/2}$ ground state are considered, under the influence of either an external magnetic or electric field.  Inelastic collisions are shown to be suppressed in the presence of modest laboratory strength magnetic and electric fields.  
 The rate of elastic collisions that rethermalize the thermal distribution, and the corresponding low rate of heating state-changing collisions, suggest that quantum degeneracy or even molecular Bose-Einstein condensation of LiO gas may be attainable, provided that the initial temperatures in the milliKelvin range are achievable.

\end{abstract}


\maketitle


\section{Introduction}

Among many successful methods for producing ultracold molecular gases \cite{Bohn17_Sci} magnetic trapping is making rapid progress.  Advances in Zeeman deceleration \cite{Vanhaecke07_PRA,Narevicius07_NJP,Momose13_PCCP,Cremers18_PRA,Akerman17_PRL} have led to trapped samples that are dense enough for molecular collisions to be observed \cite{Segev19_Nature,Plomp20_JCP}.  This work leads to exciting possibilities like magnetically controlling chemical reactivity of paramagnetic species \cite{Tscherbul09_NJP,Krems} or creating quantum degenerate molecular samples by evaporative cooling.

The experiment in Ref. \cite{Segev19_Nature} successfully measured the ratio of elastic to inelastic scattering rates for a sample of oxygen molecules, at a temperature of $\sim 800$ mK.  The observed ratio, $\gamma = K_{\rm elastic}/K_{\rm inelastic} \approx 4-8$, proves  too small by at least an order of magnitude to successfully carry off evaporative cooling, as was anticipated in earlier theoretical studies \cite{Avdeenkov01_PRA,Tscherbul09_NJP,Perez11_JCP}.  For efficient evaporative cooling, the required value of $\gamma$ is expected to be $\sim 100$ or greater.  Such a ratio has been sought in various paramagnetic species with varying degrees of success.  

In anticipation that the variety of magnetically trapped and colliding species will continue to expand, we here make an initial investigation into the collisional behavior of  the paramagnetic species LiO, which possesses a $^2\Pi_{3/2}$ electronic ground state.  This molecule also has hyperfine structure due to the nuclear spin of Li, whereby the stretched state has magnetic quantum numbers $|FM_F \rangle = |33 \rangle$.  This state is subject to two-body collisions that can transfer the molecules to lower-energy Zeeman states, or else change their parity, thus releasing energy and leading to trap loss and heating.  These are the processes that contribute to the loss rate $K_{\rm inelastic}$.  Fortunately, the molecules are not chemically reactive at ultralow temperature, as the reaction 2LiO $\rightarrow$ Li$_2$ + O$_2$ is endothermic by 9260 K.  Thus the fine-structure state changing collision are the only loss to worry about.  

Within a model of ultracold scattering in which the relevant  losses are driven primarily by the long-range electric dipole-dipole interaction between molecules, we compute $K_{\rm inelastic}$ and $K_{\rm elastic}$ for this radical.  Strikingly, over a range of modest magnetic field, $K_{\rm inelastic}$ appears to be a decreasing function of field, whereby inelastic scattering can be suppressed \cite{Ticknor05,Augustovicova18_PRA}.  Moreover, the relatively large dipole moment of LiO emphasizes the elastic scattering rate.  As a result, the ratio $\gamma$ can exceed 100 up to temperatures of hundreds of microKelvin, putting evaporative cooling potentially in reach of the new technology.

\section{Model}

The basic model for ultracold collisions of $^2\Pi$ molecules, with interaction driven by dipole-dipole forces, has been developed elsewhere \cite{Avdeenkov02,Ticknor05,Augustovicova18_PRA}.  Here we summarize the salient parts.  The two-body Hamiltonian is 
\begin{eqnarray}
H = T + V_{\rm disp} + V_{\rm d} + H_1 + H_2,
\end{eqnarray}
where $T$ is the kinetic energy, $V_{\rm disp}$ is a long-range dispersion interaction, $V_{\rm dd}$ is the dipole-dipole interaction between the molecules. The Hamiltonians of the separated molecules are given by
\begin{eqnarray}
H_i = H_{\rm rot,i} +  H_{\rm so,i} + H_{\Lambda,i} + H_{\rm hf,i} + H_{{\rm Z},i} + H_{{\rm S},i},
\end{eqnarray}
whose terms describe, respectively, the rotation, spin-orbit, Lambda-doubling, hyperfine, Zeeman, and Stark  interactions of molecule $i$.  

\subsection{The zero-field effective Hamiltonian for LiO}

 The Hamiltonian of each molecule is evaluated in a Hund's case (a) basis set.  Before incorporating the hyperfine interaction, this set has as quantum numbers the total electronic orbital and spin angular momentum projections $\Lambda$ and $\Sigma$ along the molecular axis, with sum $\Omega = \Lambda + \Sigma$; as well as the total angular momentum $J$ and its projection $M$ on the laboratory fixed axis.  For the $^2 \Pi$ states of LiO, these basis sets are denoted by the shorthand
 \begin{eqnarray}
 |\Omega \rangle | \Omega ,J,M\rangle =  |\Lambda,S,\Sigma; \Omega \rangle|\Omega,J,M\rangle.
 \end{eqnarray}
 In zero electric field the energy eigenstates are states of good parity, denoted $p=e$ or $f$, and given for the ground state by the linear combinations
 \begin{eqnarray}
 &&|^2\Pi_{|\Omega|}(e/f); JM\rangle = \\
 && \qquad \frac{1}{\sqrt{2}}\bigg( |\Omega| \rangle | \Omega| ,J,M\rangle
 \mp| -|\Omega| \rangle|-|\Omega|,J,M\rangle \bigg)\,. \nonumber 
\label{parityef}
\end{eqnarray}

In the $J=3/2$ ground state of interest here, the fine structure states are given by $^2\Pi_{1/2}$ and $^2\Pi_{3/2}$, of which $^2\Pi_{3/2}$ is the lower-lying state because the spin-orbit coupling constant for the ground vibrational state is negative, as it is for OH and SH.  Relative to this ground state, the first rotational excitation with $J=5/2$ is higher in energy by $\sim 8.7$~K, given the rotational constant  $B=1.73$\,K \cite{Yamada93}), while the first fine structure excited state is higher in energy by an energy on the scale of the fine structure constant $|A| = 111.94$~cm$^{-1}$ $\sim 160$~K \cite{Yamada93}.  For the sub-Kelvin collision energies we deal with, we will therefore ignore these excited states, and consider $J=3/2$, $|\Omega|=3/2$ to be good quantum numbers.  Given the small scale of the Lambda-doublet splitting, $\Delta_{\Lambda}\!=\,5.4\times10^{-4}$\,K \cite{Freund72},  this interaction must be incorporated, at least in low electric fields.

Each component of the doublet (\ref{parityef}) is further split by magnetic hyperfine structure, due to $^7$Li nuclear spin $I=3/2$ (the spin of the $^{16}$O nucleus is 0), into hyperfine components characterized by the total angular momentum $\vec{F} = \vec{J}+\vec{I}$. The nuclear spin states are described by coupled basis functions $|F,M_F\rangle$, defined in the usual way,
\begin{align}
\begin{split}
|\eta,&F,M_F;p\rangle \\ 
& = \!\!\sum_{M,M_I}|^2\Pi_{|\Omega|}(e/f); JM\rangle|I,M_I\rangle \langle J,M,I,M_I|F,M_F\rangle\,,
\end{split}
\end{align}
with $\eta$ denoting the other quantum numbers not given explicitly.
The hyperfine Hamiltonian is diagonal in this basis.  For the $^2\Pi_{3/2}$ state, $J = 3/2$ rotational level the corresponding energies are adopted from Refs. \cite{Freund72,Yamada93}

\subsection{Zeeman and Stark Interactions}

The Zeeman effect arises from interaction between magnetic dipoles and an external magnetic field.  For each LiO molecule the main terms are given by \cite{Brown_78}
\begin{eqnarray}
H_{\rm Z} = -\mu_{\rm B}(g_L\vec{L}\cdot\vec{B} + g_S\vec{S}\cdot\vec{B}),
\end{eqnarray}
where $\mu_{\rm B}$ is the Bohr magneton, and $g_L=1$, $g_S=2.002319$ \cite{Bruna99} are the corresponding  $g$ factors for individual type of angular momentum. 
Additional $g$-factors due to rotational Zeeman effect, the electronic spin anisotropic Zeeman effect, the nuclear spin Zeeman effect, and parity-dependent contributions for a $\Pi$ state are typically three orders of magnitude weaker and are neglected here. 
The $\vec{B}$ vector is assumed to be aligned along the laboratory $Z$-axis that defines the quantization of $M_F$.  Matrix elements of this Hamiltonian in our basis are given in Ref.~\cite{Ticknor05}.  Significantly, the Zeeman Hamiltonian is diagonal in the parity quantum number $p$.

The Stark Hamiltonian for the molecular dipole -- electric field interaction is given by 
\begin{eqnarray}
H_{\rm S} = -\vec{d}\cdot\vec{\mathcal{E}},
\end{eqnarray}
where $\vec{\mathcal{E}}$ is the electric field, which defines the space-fixed $Z$-axis in the absence of a magnetic field; and $d=6.84$~D is the electric dipole moment.  Matrix elements of this Hamiltonian are also derived elsewhere \cite{Avdeenkov02,Ticknor05,Augustovicova18_PRA}.  This interaction preserves the parity of the molecules for fields below a characteristic value $\mathcal{E}_0=5\Delta_{\Lambda}/6d \sim 2.7$ V/cm for $|M|=3/2$, while at higher fields the parity states mix, until the signed values of $\Omega$ become good quantum numbers at large  electric fields.  

\begin{figure}[h]
\includegraphics[width=0.47\textwidth]{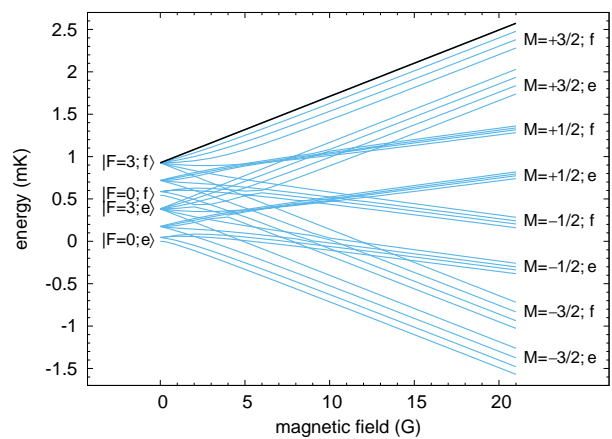}
\caption{Zeeman energies of the hyperfine  and $\Lambda$-doublet levels for $J=3/2$ of the $^2\Pi_{3/2}$ state of the LiO molecule at zero electric field. At low fields, states are indexed by total spin $F$ (not all of which are displayed, for clarity), along with the parity.  At high fields, states are indexed by the projection of the molecule's rotation, $M$, and the parity.  The hyperfine state of our interest for magnetic trapping, $|3,3;f\rangle$, is highlighted.}
\label{Zeemanenergies}
\end{figure}

\begin{figure}[h]
\includegraphics[width=0.47\textwidth]{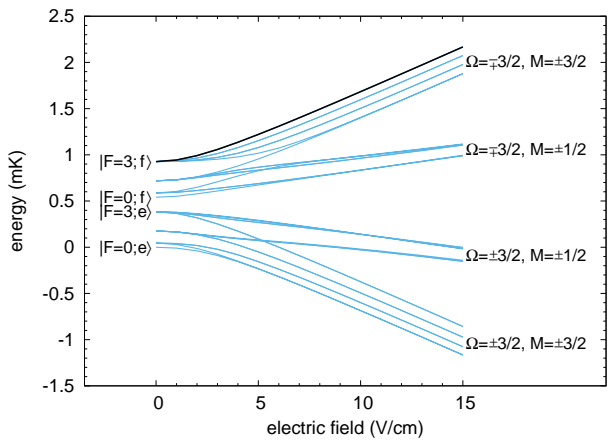}
\caption{Stark energies of the hyperfine and $\Lambda$-doublet levels for $J=3/2$ of the $^2\Pi_{3/2}$ state of the LiO molecule at zero magnetic field. At low fields,  states are indexed by total spin $F$, (not all of which are displayed, for clarity), along with the parity.  Each line is doubly degenerate for $|M_F|$. At high fields, the states are indexed by the projections $M$ and $\Omega$ of the rotational angular momentum on the laboratory and molecular axes, respectively.  The hyperfine state of our interest $|3,3;f\rangle$ is highlighted.}
\label{Starkenergies}
\end{figure}

The Zeeman and Stark energies for the $J=3/2$ ground state are shown in Figures \ref{Zeemanenergies} and \ref{Starkenergies}, respectively.  In either case when the  field is zero, the states are appropriately labeled by the total spin $F$ (not all values of $F$ are shown explicitly, to simplify the diagram).  

In magnetic fields above $B \sim10$ Gauss, (Figure \ref{Zeemanenergies}) the rotational angular momentum $M$ decouples from the nuclear spin, whereby $M$ is a  good quantum number for describing states.  Also in this instance of zero electric field, the parity label $e$ or $f$ remains a good quantum  number.  Within this classification, the additional states would be identified by the value of $M_I$ (not shown).

In electric fields above $\mathcal{E} \sim 5$ V/cm, the molecules also decouple from nuclear spins and overcome the Lambda-doublet interaction.  In this case reasonable quantum numbers are the signed values of $M$ and $\Omega$, as shown.

\subsection{Quantum scattering calculation}
At ultralow collision energies, the molecular dynamics is dominated by long-range forces. In the case of neutral diatomic molecules that possess an electric dipole moment, the most relevant interaction  between the molecules  the   dipole-dipole interaction, given by
\begin{eqnarray*}
V_{\rm d}(\vec{R}) = - \frac{ \sqrt{30} d^2 }{ 4\pi\varepsilon_0 R^3 }\!\!  \sum_{q,q_1,q_2} \!\!\!\!\!\!
&& \left( \begin{array}{ccc} 2 & 1 & 1 \\ q & -q_1 & -q_2 \end{array}\! \right)  \\
&& \times \, C_{2-q}(\theta \phi) 
C_{1q_1}({\hat n}_1) C_{1q_2}({\hat n}_2). \nonumber 
\end{eqnarray*}
Here the $C$'s are reduced spherical harmonics, $(\theta, \phi)$ are the spherical angles of the vector $\vec{R}$ joining the centers of mass of the two molecules, and ${\hat n}_i$ is the orientation of the axis of molecule $i$.  

 In addition, the leading term of the multipolar expansion comes from an attractive  dispersion interaction, which we take to be isotropic,
 \begin{eqnarray}
  V_{\rm disp} = - \frac{ C_6 }{ R^6 }.
  \end{eqnarray}
  For a highly polar molecule such as LiO, the dominant contribution to $C_6$ arises from coupling to higher-lying rotational states.  This allows us to estimate the value  $C_6= 1.99 \times 10^5$ a.u. for (LiO)$_2$.
  Details of the short-range forces are disregarded in the model, which simply declares a hard-wall boundary condition on the wave function at a radius $R=30~a_0$.  The model does not require absorbing boundary conditions at this radius, inasmuch as the LiO molecules are not chemically reactive at zero temperature.

The complete basis includes the zero-field states of each molecule, along with the partial wave state  $|L M_L \rangle$.  This basis is symmetrized with respect to the exchange of bosons, as denoted by the subscript $S$.  Basis states in general read
\begin{eqnarray}
|n \rangle = P_{12}\left\{ |\eta_1 F_1,M_{F_1};p_1\rangle|\eta_2 F_2,M_{F_2};p_2\rangle |LM_L \rangle \right\},
\label{2mol_basis}
\end{eqnarray}
where $P_{12}$ denotes the operator that exchanges the identical bosonic molecules.
For the case that the initial state consists of bosons in identical internal states, as we consider here, the partial waves are restricted to even values of $L$.  

Matrix elements of the Hamiltonian in this basis are given explicitly in \cite{Avdeenkov02,Ticknor05}.  Matrix elements of the dipole-dipole interaction have particular parity selection rules.  Specifically, in zero electric field a state where both molecules have the same initial parity $p_1=p_2=p$ (as we will assume below) are coupled directly only to those where {\it both} molecules change parity.  Vice vera, in the high electric field limit, where $\Omega$ is a good quantum number, the dipole-dipole interaction preserves the signed value of $\Omega$ \cite{Avdeenkov01_PRA}.

In the lab frame the projection of the total angular momentum $M_{\mathrm{tot}} = M_{F_1}+M_{F_2}+M_L$ is conserved throughout the collision. Considering the weak-field seeking molecular states in a magnetic or electric field and a scattering process incident on an $s$ partial wave, this projection quantum number $M_{\mathrm{tot}}$ equals 6. Calculations of collision cross section require the inclusion of partial waves up to $L=16$ for  convergence purposes. The entire basis set allowed by the $M_{\mathrm{tot}}$ conservation condition can be truncated by applying propensity rules that preferably select channels with a low value of $M_L$ as was explored in \cite{Augustovicova18_PRA}. Imposing $|\Delta M_L| \le 4$, the total number of channels is here reduced to 994.

We perform exact coupled-channel calculations for LiO-LiO scattering employing the log-derivative propagator method \cite{Johnson}. Cross sections  $\sigma$ as functions of collision energy are computed from the $S$-matrix elements for processes in which both molecules remain unchanged (elastic) or at least one molecule converts its internal state to another (inelastic).
The rate constants $K(T)$ for collisions at a given temperature $T$ are then derived from the total cross sections by averaging  the rate coefficient $v_i \sigma$ over a Maxwellian velocity distribution of initial velocities $v_i$,
assuming that the system is found in thermodynamic equilibrium. 

\section{Results and Discussion}

Considering the possibility of magnetic traps of ultracold molecules, now approaching the mK regime \cite{Segev19_Nature}, we are interested in weak-magnetic-field seeking states, such as the spin-stretched state with $|F_1M_{F1}; p_1 \rangle |F_2M_{F2}; p_2 \rangle = |33;f \rangle |33;f\rangle$.  For parity $f$, this is the state of highest energy in the ground state manifold, and is indicated by the heavy line in the Zeeman diagram of Figure \ref{Zeemanenergies}.  

Our primary goal is to assess the stability of the ultracold LiO gas against two-body inelastic collisions, while preserving a high elastic collision rate that can guarantee thermal equilibrium of the gas.  The figure of merit for calculations is then the elastic and inelastic collisions rates, and more importantly, their ratio $\gamma = K_{\rm elastic}/K_{\rm inelastic}$.  Ideally this ratio is on the order of $\gamma \sim 100$ or higher for effective evaporative cooling to occur.

\begin{figure}[h!]
\includegraphics[width=0.47\textwidth]{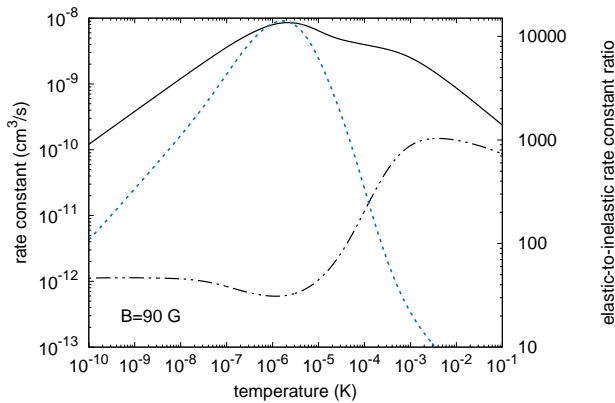}
\caption{Rate constants for elastic (solid curve) and inelastic (dash-dotted curve) scattering, along with their ratio (dotted curve, right-hand
axis) as a function of temperature at $B=90\,$G. The collision is initiated in the states $|3,3;f\rangle$ of the molecules.}
\label{rate_constants}
\end{figure}

A set of  collision  rates are  shown in Figure \ref{rate_constants} at fixed values of electric ($\mathcal{E}=0$ V/cm) and magnetic field ($B=90$ G).  At the lowest temperatures these rates exhibit the usual Wigner threshold laws, $K_{\rm elastic} \propto \sqrt{T}$, $K_{\rm inel} \propto {\rm const}$.  

Significantly, the ratio of elastic-to-inelastic collision rates $\gamma$ remains $\sim 100$ or higher over a broad temperature range, from 0.3 mK down to 1 pK, even exceeding several thousands near $\mu$Kelvin temperatures.
The computed rate constants indicate that evaporative cooling may be plausible for this species, provided that the  molecules can be initially lowered to mK temperatures.  As a much more optimistic conclusion, once the gas attains $\mu$Kelvin temperatures, the two-body inelastic rate is fairly small, meaning that the gas may be collisionally stable.  

\begin{figure}[h!]
\includegraphics[width=0.47\textwidth]{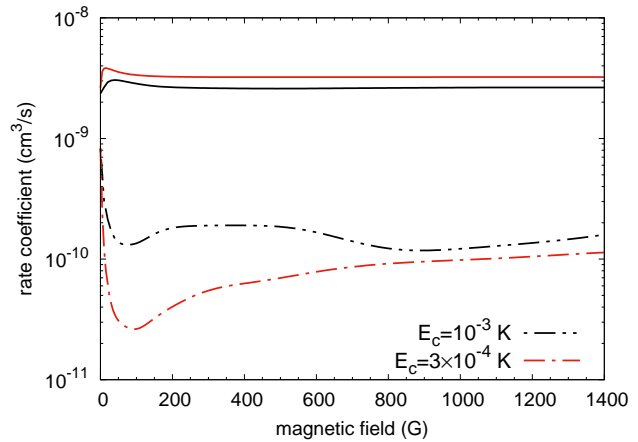}
\caption{Rate coefficients for elastic (solid curves) and inelastic (dashed curves) scattering as a function of magnetic field for incident channel of $|3,3;f\rangle$ molecular state. The collision energy is fixed at the value $E_{\rm c} = 1$ mK (black lines) and $E_{\rm c} = 0.3$ mK (red lines).
}
\label{rateB10-3}
\end{figure}

At collision energies of about mK the ratio of elastic over inelastic processes not only drops below 100, it also  and becomes rather sensitive to the applied magnetic  field as shown in Figure \ref{rateB10-3}. The inelastic rate is seen to drop rapidly for small fields, up to about $B=10$ Gauss, at which point the Zeeman interaction dominates over the hyperfine structure.  At low fields, a number of hyperfine levels are roughly equally likely to be populated after a collision.  However, at higher fields we find that the collisions are subject to more restrictive propensity rules.  Indeed, for fields above $\sim 100$ Gauss, the dominant loss channels appear to be those where the final parity has changed from $f$ to $e$ for both molecules (a consequence of the channel coupling of the dipole-dipole interaction), while the lab-frame projection of spin, $M$, is unchanged.  There is additional, minor, inelastic scattering to channels with $e$ parity and small changes of $M$.

Another feature of the inelastic rate in Figure \ref{rateB10-3} is a minimum near $B=90$ Gauss, which indeed is what prompted us to consider collision rates at this field in Figure \ref{rate_constants}.  This appears to be not fundamental, but somewhat fortuitous.  The dominant loss channels happen to have an interference minimum under the circumstances shown.  While such minima can be described  in the distorted wave Born approximation \cite{Ticknor05}, they cannot be predicted without detailed knowledge of short-range phase shifts.  Finding such a minimum empirically would of course be useful for minimizing the inelastic rates.  

\begin{figure}[h!]
\includegraphics[width=0.47\textwidth]{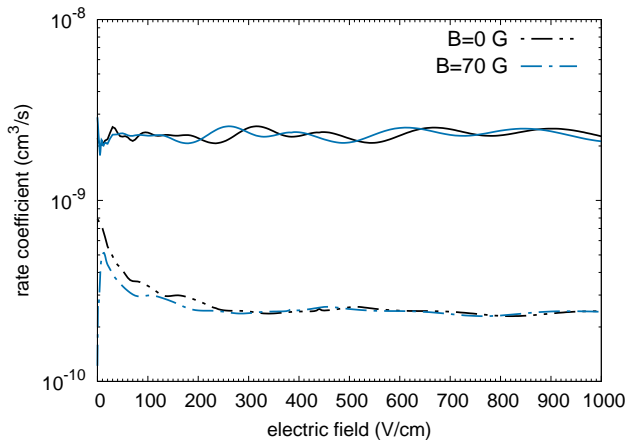}
\caption{Rate coefficients for elastic (solid curves) and inelastic (dashed curves) scattering as a function of electric field at $B=0$\,G (black curve) and $B=70$\,G (blue curve). This magnetic field is colinear with the electric field.  The collision energy is fixed at the value $E_{\rm c} = 1$ mK for incident channel of $|3,3;f\rangle$ molecular state.}
\label{rateE10-3}
\end{figure}

A similar overall behavior occurs when an electric field is applied, irrespective of the application of a small magnetic field (Figure \ref{rateE10-3}).  For very small fields, the inelastic rates rise when $\mathcal{E} < \mathcal{E}_0$ \cite{Avdeenkov01_PRA}.  Then there is a drop as propensity rules favor a small number of exit channels, those with $\Omega$ conserved and $\Delta M \le 2$.
  Fig. \ref{rateE10-3} indicates that small electric field of several hundred V/cm can act to suppress inelastic collisions at $E_{\rm c} = 1$ mK, but the suppression does not come particularly close to achieving the desired goal of $\gamma  \sim 100$.

\section{Conclusion}
In this paper we studied the ultracold collisions of polar LiO molecules in their ground electronic rovibrational state, focusing on the weak-field-seeking state in an external magnetic or electric field.
Relatively weak fields may have a profound influence on the collision dynamics when applied separately.
We have shown that this molecular species possess a sudden drop in inelastic collisions at values of the applied magnetic field that are even in the order of tens of G, thus improving the elastic to inelastic rates to their favorable ratios, provided that the temperatures of the trapped gas are achievable by laser cooling.
The electric field can assist in increasing this ratio of efficiency, however it does not sufficiently control the suppression of inelastic rates.
The rate of elastic collisions that rethermalize the thermal distribution, accompanied by a low rate of heating state-changing collisions, indicate that quantum degeneracy or even molecular Bose-Einstein condensation of LiO gas may be feasible.

\section*{Acknowledgements}  
This material is based upon work supported by the National Science Foundation under Grant Number PHY 1734006 and Grant Number PHY 1806971.  L.D.A. acknowledges the financial support of the Czech Science Foundation (Grant No. 18-00918S).

\bibliography{Bibliography}

\end{document}